\pdfoutput=1
\documentclass[12pt,letterpaper]{JHEP3}
\usepackage{graphics}
\usepackage{epsfig}
\usepackage{epstopdf}

\usepackage{graphics,epsfig}
\usepackage{graphicx}% Include figure files
\usepackage{bm}% bold math
\usepackage{amsmath}
\usepackage{mathrsfs}

\newcommand{\be}{\begin{equation}}
\newcommand{\ee}{\end{equation}}
\newcommand{\beq}{\begin{eqnarray}}
\newcommand{\eeq}{\end{eqnarray}}

\def\nue{\mathrel{{\nu_e}}}

\def\nux{\mathrel{{\nu_x}}}

\def\barnue{\mathrel{{\bar \nu}_e}}

\def \lta {\mathrel{\vcenter{\hbox{$<$}\nointerlineskip\hbox{$\sim$}}}}
\def \gta {\mathrel{\vcenter{\hbox{$>$}\nointerlineskip\hbox{$\sim$}}}}

\def\t13{\mathrel{{\theta_{13}}}}
\def\y12{\mathrel{{\tan^2 \theta_{12}}}}
\def\c2{\mathrel{{\chi^2 }}}

\newcommand{\new}[1]{{ \textit{#1} }}

\newcommand{\n}{neutrino}
\newcommand{\ns}{neutrinos}

\newcommand{\sn}{supernova}
\newcommand{\sne}{supernovae}

\title{An ``archaeological"  quest for galactic supernova neutrinos}

\author{Rimantas Lazauskas\\
IPHC, IN2P3-CNRS/Universit\'e Louis Pasteur BP 28, F-67037 Strasbourg Cedex
2, France}

\author{Cecilia Lunardini\\
Arizona State University, Tempe, AZ 85287-1504\\
RIKEN BNL Research Center, Brookhaven National Laboratory, Upton, NY 11973}

\author{Cristina Volpe\\
Institut de Physique Nucl\'eaire, F-91406 Orsay cedex, France}

\abstract{We explore the possibility to observe the effects of electron neutrinos
from past   galactic  \sne, through a geochemical measurement of the amount of Technetium 97 produced by
neutrino-induced reactions
in a Molybdenum ore. The calculations we present take into account the recent advances in our knowledge of
neutrino interactions, of  \n\ oscillations inside a supernova, of the solar neutrino flux at Earth and of possible failed \sne.
The predicted Technetium 97 abundance is of the order of $10^7$ atoms per 10 kilotons of ore, which is   close
to the current geochemical experimental sensitivity.  Of this, $\sim 10-20\%$ is from \sne. Considering the comparable size of uncertainties,  more precision in the modeling of  neutrino fluxes as well
as of neutrino cross sections is required for a meaningful measurement.
}

\begin{document}

\section{Introduction}
\label{intro}
The question of the rate of core-collapse supernovae  in our galaxy is as interesting -- in connection with the history of star formation,
 the origin of elements and climate evolution -- as it is  difficult.

Next to more traditional approaches -- based on astronomical searches,
observation of radioactive decays and historical records --  an interesting
 combination of neutrino physics, nuclear physics and
geochemistry offers a way to count past supernovae and learn about their
neutrino output.  The
idea, first proposed in 1988 by Haxton and Johnson (HJ)
\cite{Haxton:1987bf}, relies on the fact that energetic electron neutrinos
can produce Technetium 97 ($^{97}{\rm Tc}$) from natural Molybdenum ($^{98}{\rm
Mo}$ and $^{97}{\rm Mo}$) via the reactions $^{98}$Mo$(\nu_e,e^-n)^{97}$Tc and the $^{97}$Mo$(\nu_e,e^-)^{97}$Tc.
The lifetime  of $^{97}{\rm Tc}$ is $2.6 \cdot 10^6$
years, which eliminates the possibility of the $^{97}{\rm Tc}$ found on earth
to be primordial. Therefore presence of $^{97}{\rm Tc}$ in a Molybdenum ore
is a result of either neutrino or cosmic-ray reactions. 
Any excess of $^{97}{\rm Tc}$, over the expected contribution of solar \ns\ and of cosmic-rays,
is an historical imprint of nearby supernova explosions that
happened before any human technology was there to study them! This suggests the possibility for an ``archaeological" quest of the neutrinos from past galactic supernovae, through a geochemical measurement. 

In spite of the challenges and uncertainties involved, Haxton and Johnson came
to positive conclusions: the neutrino-induced production of $^{97}{\rm Tc}$ in
Molybdenum rocks, such as the Henderson mine in Colorado, was estimated to happen at an average of  $1.57 \cdot 10^{-36}$ captures per second per target
atom,  about $40\%$ of the background due to solar neutrinos, and corresponding to a detectable abundance of $^{97}{\rm Tc}$ at the present time.
It was understood that the background due to cosmic rays could, in
principle, be reduced to acceptable levels by excavating deeply enough
and/or measuring the $^{97}{\rm Tc}$ attenuation rate as a function of the excavation depth.

In 1988 the idea was ahead of its time, in a field mostly focused on the disappearing
solar and atmospheric neutrinos. Therefore it remained restricted to a small community \cite{Nguyen:2005dq}.  Initial experimental efforts, led by K. Wolfsberg (LANL), lacked the funds necessary to overcome technical difficulties, and were soon abandoned
 \cite{haxtonprivate,NYTimes,amm}.
 
Today the field is more mature: after solving the solar and atmospheric anomalies and discovering neutrino oscillations, 
research has focused more on supernova neutrinos, whereas supernova models
have become sufficiently predictive.
a geochemical measurement of $^{97}{\rm Tc}$ production could constrain the
galactic supernova rate (the extragalactic contribution is negligible), for a given model of neutrino spectra and
luminosities from an individual supernova, without the  extinction effects typical of astronomy.
Notably, even dark neutrino sources  -- e.g., those that give no luminous signal due to rapid formation of a black hole -- would be counted.
The capability to trace back to ancient \sne\ in our galaxy  would be an unique complement to the activity of real time, large scale 
liquid detectors. With the masses of 0.1-1 Mt envisioned for the near future \cite{Barger:2007yw,Autiero:2007zj},  these detectors should see the diffuse flux of \ns\ that continuously reaches the Earth from all the \sne\ of the universe and that has a substantial cosmological component  (see e.g. \cite{Ando:2004hc}). 
Finally, the $^{97}{\rm Tc}$ data could be the only one available on
\emph{electron neutrinos} from supernovae! They would add complementary information compared to the $\barnue$ data from SN1987A \cite{Hirata:1987hu,Bionta:1987qt,Alekseev:1988gp}.   It is fascinating that this information, so important to
have a complete picture of  core-collapse,
is there, buried in Molybdenum rocks, only waiting to be
deciphered.

 In the present paper we perform detailed calculations of the expected amount of Technetium 97 produced by the neutrino-induced reactions in Molybednum ore, due to both solar and supernova neutrinos. Our predictions are based on the \new{best available} results on the solar neutrino fluxes and on the recent developments in the study of \n\ oscillations inside the star.
To estimate uncertainties, we also consider a range of values for the galactic supernova rate as well as for the neutrino fluxes at the neutrino-sphere, in accordance with supernova modeling.
The structure of the paper is as follows. In Section II and III we
give generalities on the \sn\ galactic rate and on the solar and supernova neutrino fluxes.
Section IV describes
calculations of the neutrino-nucleus cross sections. The results are summarized in Section V, while conclusions are in Section VI.

\section{Supernovae in our galaxy: rates and distances}

The rate of core collapse \sne\ (SNe) in the Milky Way is known only up to its order of magnitude.  The rate $R_{sn}\simeq 0.09~{\rm yr^{-1}}$ used by HJ, is practically excluded by the lower values measured recently  (table \ref{tab:rate}).
Scaling from external galaxies and gamma rays from galactic $^{26}$Al independently favor a rate of $R_{sn}\simeq 0.02~{\rm yr^{-1}}$ as central value, and up to $R_{sn}\simeq 0.05~{\rm yr^{-1}}$ at $3 \sigma$.   These are consistent with the much more uncertain estimate from historical records of galactic SNe, which favors a higher rate.  The upper limit from the non-observation of \sn\ \n\ bursts is rather loose, but still of interest for its being extinction-free and sensitive to possible failed \sne, as mentioned in sec. \ref{intro}.  

\begin{table}[hbt]
 \caption{\label{tab:rate}Estimated rate of galactic core-collapse
 SNe per century, from \cite{Raffelt:2007nv}, with permission of the author.}
 \centering
 \begin{tabular}{llll}
 \hline
 Method&Rate&Authors&Refs.\\
 \hline
 Scaling from external galaxies&$2.5\pm0.9$
 &van den Bergh  &
 \cite{vandenBergh:1994,Diehl:2006cf}\\
 &&\& McClure (1994)\\
 &$1.8\pm1.2$&Cappellaro \&\ Turatto&
 \cite{Cappellaro:1999qy,Cappellaro:2000ez}\\
 &&(2000)\\
 Gamma-rays from galactic $^{26}$Al&$1.9\pm1.1$
 &Diehl et al.\ (2006)&
 \cite{Diehl:2006cf}\\
 Historical galactic SNe (all types)&$5.7\pm1.7$&
 Strom (1994)&\cite{Strom:1994}\\
 &$3.9\pm1.7$&
 Tammann et al.\ (1994)&\cite{Tammann:1994ev}\\
 No neutrino burst in 25 years$^{a}$&${}<9.2$ &
 Alekseev \& Alekseeva&\cite{Alekseev:2002ji}\\
 &(90\% CL)&(2002)\\
 \hline
 \multicolumn{4}{l}{$^a$The limit of
 Ref.~\cite{Alekseev:2002ji} is scaled to 25~years of neutrino sky coverage.}
 \end{tabular}
\end{table}

Here we keep $R_{sn}$ as a free normalization; results for transition rates will be given for $R_{sn}\simeq 0.01~{\rm yr^{-1}}$, as they can be easily rescaled for other values.  The final results for the $^{97}{\rm Tc}$ abundance (table \ref{table:abundances}) refer to the more realistic $R_{sn}\simeq 0.03~{\rm yr^{-1}}$, for easier comparison with previous literature on \sn\ \ns\ where this rate is commonly used. 

To describe the spatial distribution of supernovae in our galaxy we use  cylindric coordinates $(r,z,\theta)$ centered at the galactic center, which is at $d_{\odot}\simeq 8.5$ kpc of distance from us.  We assume a uniform distribution in $\theta$ and  adopt the following function for the distribution in $r$ and $z$:
\beq
{n}_{\rm cc}(r)\propto r^\xi\exp(-r/u)
\times \left[  0.79~  e^{  - \left( {z/ 212
\ {\rm pc}} \right)^2 } + 0.21~ e^{ - \left( {z/
636 \ {\rm pc}} \right)^2 } \right]~,
\label{eq:distributioncyl}
\eeq
which fits observations of neutron stars with the parameters $\xi=4$ and $u=1.25~{\rm kpc}$
~\cite{Yusifov:2004fr,Ferriere:2001rg,Lorimer:2003qc}, and is illustrated in more detail in ref. \cite{Mirizzi:2006xx}.

\begin{figure}[hbt]
\centering
\includegraphics[width=0.9\textwidth]{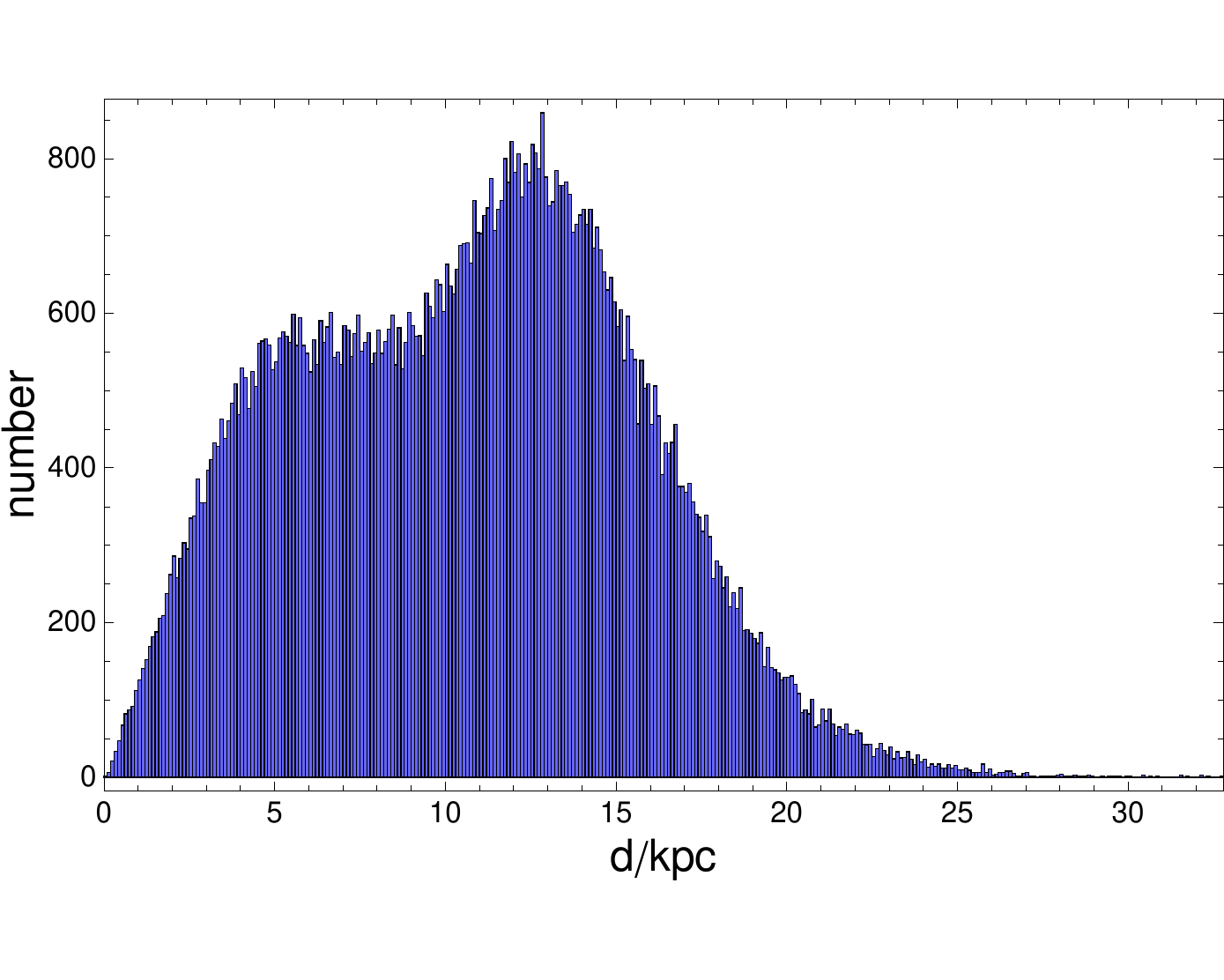}
\caption{\label{fig:distribution} The distribution with the distance from Earth of a Monte Carlo-generated population of $10^5$ \sne. The distribution in  Eq.~(\protect\ref{eq:distributioncyl}) was used, with   $\xi=4$ and $u=1.25~{\rm kpc}$.}
\end{figure}
With this, we calculate the  distribution of \sne\ with the distance from Earth, using both numerical integration and the Monte Carlo Method (fig. \ref{fig:distribution}).  We find that the neutrino flux at Earth corresponds to the flux one would obtain if all the \sne\ were at the effective distance $d_{eff}=4.26~{\rm kpc}$, which we use throughout the paper.  With the alternate set of parameters given in \cite{Mirizzi:2006xx} ($\xi=2.35$, $u=1.528~{\rm kpc}$) we get a value only slightly different, $d_{eff}=4.6~{\rm kpc}$.

The Monte Carlo approach was used to investigate possible variations in the results due to one single supernova very close to the Earth, $d \lta 100$ pc or so.   Our results agree with those in HJ on how such variations can be safely neglected: only very few ($\sim 2$ in $10^5$) events satisfy this condition, therefore the uncertainty due to the  fluctuation of \sn\ positions in the galaxy turns out to be  small compared to other uncertainties of different nature (total \n\ luminosity per \sn, etc.). 

We have not investigated effects of local deviations from the spatial distribution (\ref{eq:distributioncyl}).  Nguyen and Johnson have studied this possibility \cite{Nguyen:2005dq}, pointing out that a particularly high   \sn\ rate in the Sco-Cen region of the galaxy might result in an increase of the $^{98}$Mo$(\nu_e,e^-n)^{97}$Tc production rate, to a level equivalent to  the whole galactic \sn\ \n\ production with $R_{sn}=0.01$ ${\rm yr^{-1}}$ rate. The full investigation of this possibility is beyond the scope of our paper; still one can effectively include it here as a further uncertainty in the rate $R_{sn}$, which is a free parameter for us.

%%%%%%%%%%%%%%%%%%%%%%%%%%%%%%%%%%%%%%%%%%%%%
\section{Solar and supernova neutrino fluxes}
\label{sec:fluxes}
%%%%%%%%%%%%

\subsection{Neutrinos from  supernovae}

A core collapse supernova is an extremely powerful
neutrino source. It releases about $3 \cdot  10^{53}$ ergs of energy
within $\sim$10 seconds in neutrinos and antineutrinos of
all flavors before the star explodes, leaving behind a neutron star or, in rare cases, a black hole.

The spectrum of the neutrinos of a given flavor $w$ at production in the star is approximately
thermal, and commonly described by the form \cite{Keil:2002in}:
\be
F^0_w = \frac{(1+\alpha_w)^{1+\alpha_w} L_w}{\Gamma(1 +\alpha_w) E^2_{0w}} \left( \frac{E}{E_{0w}}  \right)^{\alpha_w} e^{-(1 +\alpha_w)E/E_{0w}}~,
\label{spectrumprod}
\ee with average energies $E_{0 w}$ in the range of $10 - 20$ MeV and
with the muon and tau species (collectively called $\nux$ from here
on) being harder than the electron ones due to their weaker (neutral
current only) coupling to matter.  $L_w $ is the (time-integrated)
luminosity in each flavor, $L_w \sim 5 \cdot 10^{52}~{\rm ergs}$;
$\alpha_w$ is a parameter describing the shape of the spectrum,
$\alpha_w \simeq 2 - 5$ \cite{Keil:2002in}, with larger $\alpha_w$ corresponding
to narrower spectrum.

For the purpose of studying the dependence of
our results on the original \n\ fluxes, we choose three sets of
parameters (Table \ref{table:inputs}) that are inspired by current
numerical calculations \cite{Totani:1997vj,Keil:2002in,Thompson:2002mw} (see also \cite{Ando:2004hc}  for a summary of those)
and that can be considered as representative of natural and extreme
cases.  The cases of an especially high or low $\nue$ flux at Earth
are modeled mostly by varying the average energy and luminosity of the
original $\nux$ flux.  This is justified by the fact that this flux is
responsible for at least $\sim 70\%$ of the $\nue$ flux in a detector,
as will be clarified below. 
\begin{table}[h!]
\begin{center}
\begin{tabular}{|c|c|c|c|}
\hline
  & best & natural & worst \\
\hline
\hline
$(E_e,E_x)/{\rm MeV} $  & (13,22) & (12,18) & (10,16) \\
\hline
$(L_e,L_x)/L_0 $ & (1,2) & (1,1) &  (1, 0.5)\\
\hline
$(\alpha_e,\alpha_x) $ & (3.5,2.5) &  (3.5,2.5) & (3.5,2.5) \\
\hline
\hline
\end{tabular}
\end{center}
\par
\vskip 0.5cm
\caption{The three sets of input parameters for the neutrino fluxes used in this paper.  Here $L_0=0.5 \cdot 10^{53}~{\rm ergs}$.}
\label{table:inputs}
\end{table}

As they propagate from the production point to a detection point on Earth,  the
neutrinos undergo  flavor conversion (oscillations).
Therefore, up to a geometric factor $1/4 \pi d^2$, with $d$ the distance star-Earth, the flux of $\nue$ in a detector at Earth depends on the fluxes at production as:
\be
F_e = p F^0_e + (1-p) F^0_x  ~,
\label{spectrumearth}
\ee with $p$ the $\nue$ survival probability.  For a given density
profile of the star $p$ is determined by number of physical phenomena,
namely: (i) \n\ -\n\ coherent scattering, which induces complicated non-linear effects \cite{Samuel:1993uw,Pastor:2001iu,Balantekin:2004ug,Duan:2005cp,Hannestad:2006nj,Fogli:2007bk,Raffelt:2007xt,Dasgupta:2007ws,EstebanPretel:2007yq,Gava:2008rp} (ii)
two MSW resonances produced by \n-electron scattering \cite{Mikheev:1986if,Dighe:1999bi,Lunardini:2003eh} and
(iii) oscillations inside the Earth \cite{Dighe:1999bi,Lunardini:2001hc,Dasgupta:2008my}. 
As a result  $p$ is a function of the \n\ energy, the mixing angle $\t13$ and
the mass hierarchy (ordering) of the neutrino mass spectrum, i.e.,
the sign of the mass squared splitting $\Delta m^2_{31}$ (see e.g.,
\cite{Dighe:1999bi}).  For the \emph{ normal} mass hierarchy, $\Delta
m^2_{31}>0$, $p$ varies in the interval
\be
p= 0 - \sin^2 \theta_{12}
\simeq 0 - 0.31 ~,
\label{prob}
\ee
 as $\sin^2 \t13$ increases in the interval $\sin^2
\t13 \simeq 10^{-5} - 10^{-2}$ \cite{Dighe:1999bi,Lunardini:2003eh}. Here $\theta_{12}\simeq 34^\circ$ is the
``solar'' mixing angle (see e.g. \cite{Aharmim:2008kc}).  For the extreme values of $\theta_{13}$, the dependence of the survival probability on the energy is negligible in first approximation; oscillations in the Earth are also at the level of few per cent when averaged over the different arrival directions of the \ns\ \cite{Lunardini:2005jf}, therefore they will be neglected here.

For  $\Delta m^2_{31}<0$ the calculation of $p$ is complicated by
the effects of \n\ -\n\ scattering, which could be strong  few seconds after the start of the \n\ burst \cite{EstebanPretel:2008ni} and induce a swap of the $\nue$ and $\nux$
fluxes above a critical energy $E_c \sim 6-9$ MeV \cite{Duan:2005cp,Hannestad:2006nj,Fogli:2007bk,Raffelt:2007xt,Dasgupta:2007ws,EstebanPretel:2007yq,Gava:2008rp}.  Still,
the survival probability remains within the interval in
Eq. (\ref{prob}) at all times (see e.g. \cite{Dasgupta:2007ws}), so one can always define a
time averaged survival probability with value in this interval.

For our purposes, it is sufficient to consider $p$ as energy independent, as $E_c$ is typically below threshold of the cross sections of interest (see sec. \ref{xsec}), and take the two extreme cases of total ($p=0$) and minimal ($p=\sin^2 \theta_{12}$) flux permutation. This is adequate to show the extent of variation of the $^{97}{\rm Tc}$ abundance  with the conversion pattern.

\begin{figure}[h!]
\centering
\includegraphics[width=0.9\textwidth]{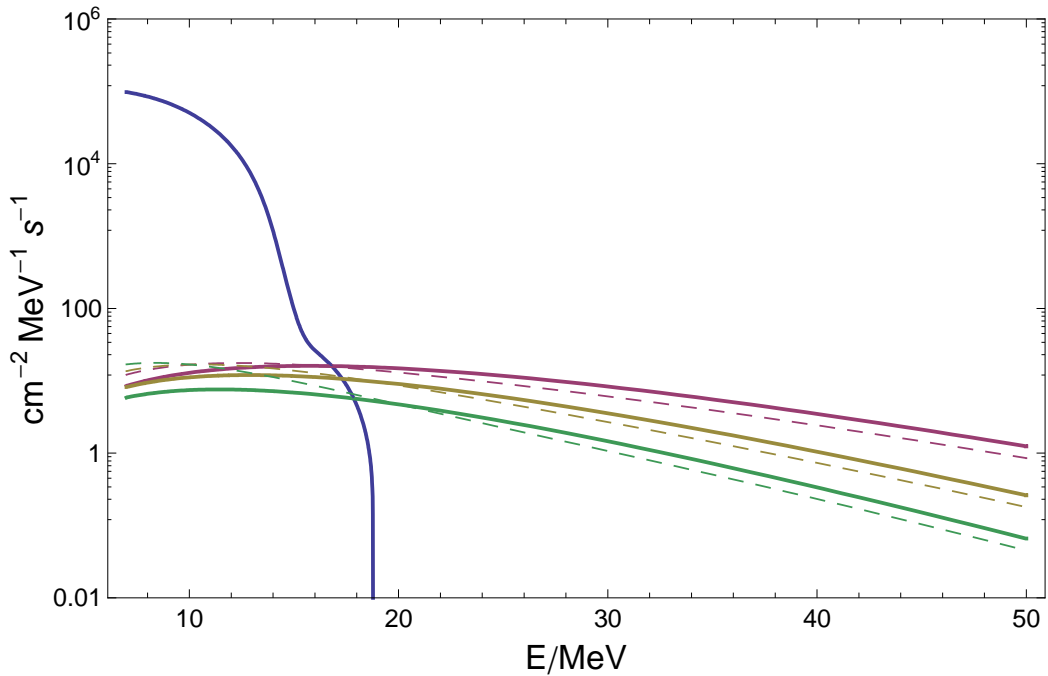}
\caption{\label{fig:fluxes}  Time averaged  solar (leftmost curve) and \sn\ (lines extending to the right) $\nue$ fluxes at Earth, inclusive of oscillation effects.    The solid curves correspond, from upper to lower, to the best, natural and worst scenarios in Table \protect\ref{table:inputs} and complete flavor permutation ($p=0$, see Eqs. (\protect\ref{spectrumearth}) and (\protect\ref{prob})).  The dashed ones refer to the same scenarios with the minimal permutation ($p\simeq 0.31$). A rate $R_{sn}=10^{-2}~{\rm yr^{-1}}$ was used.  }
\end{figure}
Fig. \ref{fig:fluxes} shows the time averaged \sn\ $\nue$ flux at Earth:
\be
F^{earth}_{e} = R_{sn} F_{e}/(4 \pi d^2_{eff})~.
\label{fluxearth}
\ee
We have plotted the six fluxes at Earth
obtained with the two extreme values of $p$ and the three scenarios in Table \ref{table:inputs}.  They are averaged over time for the purpose of comparison with the solar \n\ flux\footnote{Using a time averaged flux might be somewhat confusing, because it might suggest the idea of a flux that is continous in time.  In constrast with the solar one, the \sn\ \n\ flux we are considering is far from continous: it consists of $\sim 10$ s bursts that reach the Earth a few times per century and whose effects on the $^{97}{\rm Tc}$ production accumulate over several millions of years. }.

All that was discussed so far refers to the most common scenario of
core collapse \sn, the one that leads to a successful explosion and
the formation of a neutron star.  Stars of mass above 25-40 $M_\odot$
($M_\odot=1.99 \cdot 10^{30}$ Kg being the mass of the sun), a $10-20\%$ fraction of all
collapse candidates, could either explode and form a black hole by
fallback or collapse into a black hole directly, with no explosion (see e.g. \cite{Woosley:2002zz} and references therein).  The \n\
emission in the latter case has been studied recently \cite{Sumiyoshi:2006id,Sumiyoshi:2007pp,Fischer:2007zzb,Sumiyoshi:2008zw,Nakazato:2008vj} and
found to be characterized by a much shorter burst (${\mathcal O}(1) $
s or less) with luminosity and average energies higher than the
neutron-star forming case, especially in $\nue$ and $\barnue$. Higher
and more energetic fluxes have been found \cite{Sumiyoshi:2006id,Sumiyoshi:2007pp,Fischer:2007zzb,Sumiyoshi:2008zw} for the stiffer
equation of state (EoS) by Shen et al. (S) \cite{shenetal} compared to the softer one by
Lattimer and Swesty (LS) \cite{lattimer91generalized}.
\begin{figure}[h!]
\centering
\includegraphics[width=0.9\textwidth]{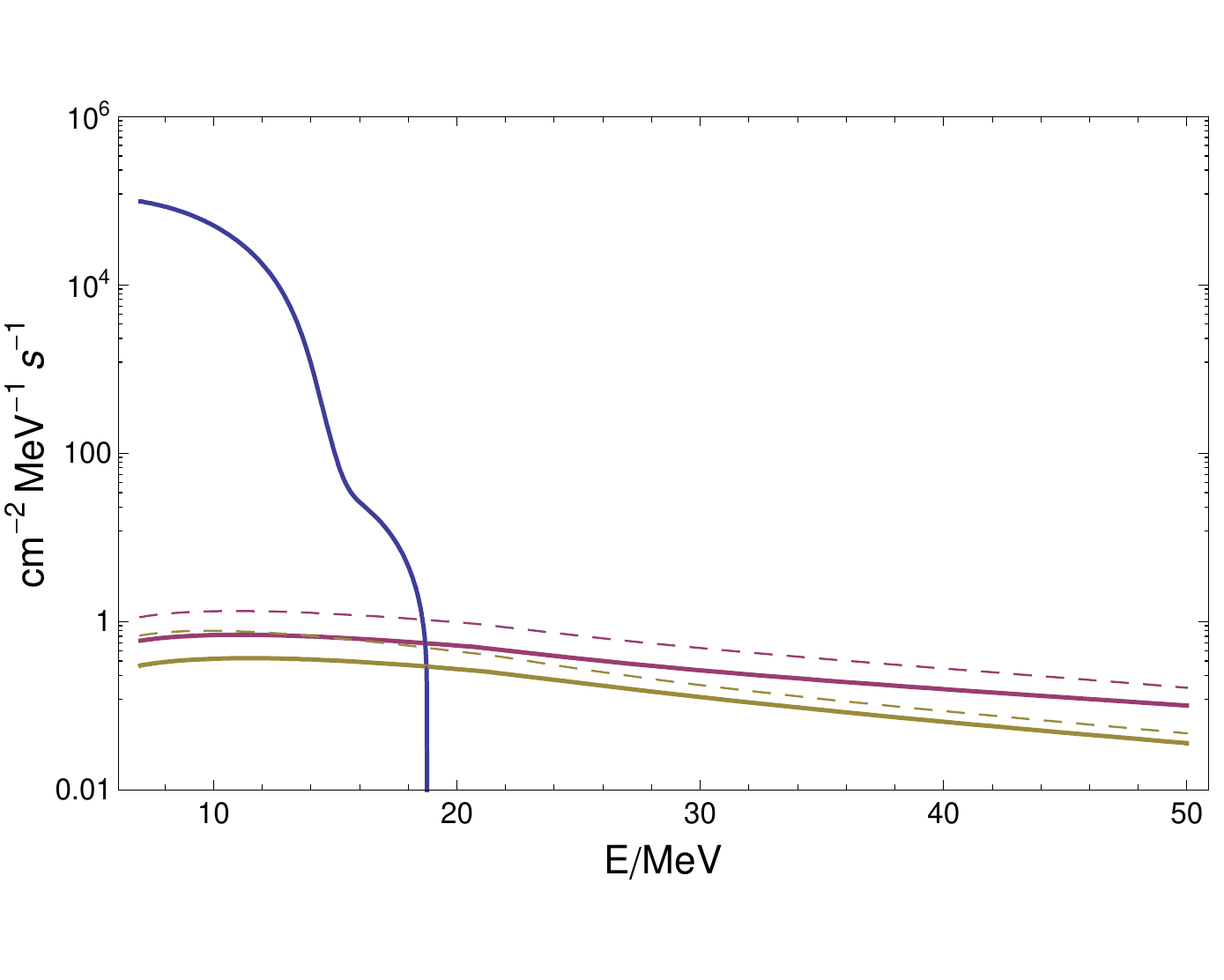}
\caption{\label{fig:fluxesbh} The same of fig. \protect\ref{fig:fluxes} for the flux of failed supernovae, obtained using the fluxes in fig. 5 of \protect\cite{Nakazato:2008vj}. Of the two solid curves, the upper (lower) is for the Shen  (Lattimer-Swesty) EoS. The same holds for the dashed lines. We took a rate for failed collapse of $R_{bh}=10^{-3}~{\rm yr^{-1}}$.}
\end{figure}

Here we model the contribution of these \emph{ failed supernovae} to
the time averaged \sn\ $\nue$ flux (shown in fig. \ref{fig:fluxesbh}) using the original fluxes presented
in ref. \cite{Nakazato:2008vj} (fig. 5 there).
 Those have the character of
examples, only roughly representing the possible variations that one
can have with the EoS and progenitor model, for which a comprehensive study still lacks. They are characterized by the parameters $E_{0x}\simeq 24$ MeV, $L_x\simeq 0.45 \cdot 10^{53}$ ergs, $E_{0e}\simeq 20.7$ MeV, $L_e \simeq 1.4\cdot 10^{53}$ ergs for the S EoS and by $E_{0x}\simeq 22$ MeV, $L_x\simeq 0.2 \cdot 10^{53}$ ergs, $E_{0e}\simeq 17$ MeV, $L_e \simeq 0.54 \cdot 10^{53}$ ergs  for the LS one. For the spectral shapes, we checked that Eq. (\ref{spectrumprod}) is a reasonable approximation; still we chose to interpolate the numerically calculated points in \cite{Nakazato:2008vj} with a polynome of order 4 in the plane $\log(E)-\log(F^0_w)$. Results have only minor differences compared to the more conservative linear fit used in \cite{Nakazato:2008vj}.

In absence of indications of the opposite, we assumed that black hole forming collapses have the same spatial distribution, Eq. (\ref{eq:distributioncyl}), as the neutron star forming ones, and that the estimated ratio of rates for the two types of collapses, $R_{bh}/R_{sn}\sim 0.1 - 0.2$ is valid for our galaxy.
We have also considered the conversion effects to be the same for the two types, Eqs. (\ref{spectrumearth}) and (\ref{prob}).  While this is not the case for every value of $\t13$, it is true for the extreme values of $\t13$ and of $p$, which are of the most interest here \cite{Nakazato:2008vj}.
Fig. \ref{fig:fluxesbh} reveals that the contribution of failed \sne\ is at the level of tens of per cent of the total flux at $E \sim 20 -30$ MeV, and can exceed the flux from regular supernovae above 50-60 MeV depending on the parameters.  Notice that for failed \sne\ the $\nue$ flux is smaller for total flavor permutation $p=0$, because in this case the flux of $\nue$s at Earth originates from the less luminous original $\nux$ flux.

%\subsection{Failed supernovae: direct collapse into a black hole}

\subsection{Solar neutrinos}
Solar \ns\  are the only significant background that can not be eliminated.
In the energy range of interest here, $\nue$s from the $^8{\rm B}$  and $hep$ reactions of the fusion chain are relevant.  The former component extends to 16.6 MeV energy, while $hep$ \ns\ reach 18.8 MeV.
We describe these fluxes following  the Bahcall-Serenelli-Basu BS05(PO) solar model \cite{Bahcall:2004pz}. There, the  $^8{\rm  B}$ component is normalized to $\Phi_{B}=5.69\cdot 10^6~{\rm cm^{-2} s^{-1}}$, and the $hep$ one  to
$\Phi_{hep}=7.93\cdot 10^3~ {\rm cm^{-2} s^{-1}} $ (unoscillated
values).
The solar neutrino survival probability as a function of energy was taken
as the best fit in \cite{Fogli:2007tx} (fig. 2 of this reference), corresponding to the LMA-MSW oscillation parameters $\Delta m^2_{21}=7.92 \cdot 10^{-5}~{\rm eV^2} $ and $\sin^2 \theta_{12}=0.314$.
With these, the \emph{ oscillated} fluxes at Earth are $ \Phi^{osc}_{B}=2.17 \cdot 10^6~{\rm cm^{-2} s^{-1}}$, and  $\Phi^{osc}_{hep}=2.81   \cdot 10^3~ {\rm cm^{-2} s^{-1}} $.  The total of the two, differential in energy, is shown in figs. \ref{fig:fluxes} and \ref{fig:fluxesbh}.  

Notice that the oscillated fluxes at Earth used here  differ substantially from those used by Haxton \& Johnson: $ \Phi^{HJ}_{B}=1.1 \cdot 10^6~{\rm cm^{-2} s^{-1}}$, and  $\Phi^{HJ}_{hep}=1.6  \cdot 10^4~ {\rm cm^{-2} s^{-1}} $.    This difference is a result of the two decades of progress in solar \n\ research, and is one of the major novelties of our work.

A comment on uncertainties on the solar \n\ flux is due.  Errors on the measured  $\Delta m^2_{21}$ and $\sin^2 \theta_{12}$  are of $\sim 3\%$ and $\sim 10\%$ (at $1\sigma$) respectively \cite{Aharmim:2008kc}; corresponding to an error of the order of $\sim 10\%$ on the  solar \n\ survival probability.
The solar model itself has uncertainties, e.g. those associated to using different inputs for the solar opacities \cite{Bahcall:2004pz}, which are at the level of ${\mathcal O}(0.1)\%$.
Here we do not include all these uncertainties explicitly, because they do not impact our conclusions significantly; moreover, we note that for our purposes the oscillated fluxes $\Phi^{osc}_{hep}$ and $ \Phi^{osc}_{B}$ are the only relevant quantities (assuming their spectra as known).  These are measurable very precisely, in principle.
The SNO experiment has already measured $ \Phi^{osc}_{B}$ (the part of the flux above its threshold of $\sim 5$ MeV) with precision of $\sim$ 3-5\% (statistical and systematic, 1$\sigma$) \cite{Aharmim:2008kc}, while a measurement of $hep$ \ns\ requires to wait for the next phase of solar detectors \cite{robertson08}.
The importance of a precise measurement of the solar \n\ fluxes in view of a possible ${\rm ^{97}Tc}$  experiment will be discussed briefly in sec. \ref{sec:results}.

%%%%%%%%%%%%%%%%%%%%%%%%%%%%%%%%%%%%%%%%%%%%%%%%%%
\section{Neutrino-induced cross sections}
\label{xsec}

Predictions of neutrino-nucleus cross sections in the solar and supernova energy range require
a precise knowledge of the transitions to the low lying states and to the resonance region, well
above the particle emission thresholds.
The latter include both allowed transitions associated
to the Gamow-Teller and Isobaric Analogue State - and the forbidden ones, such as the spin-dipole and higher multipoles. The particular reaction cross sections of interest
are $^{98}$Mo$(\nu_e,e^-n)^{97}$Tc and $^{97}$Mo$(\nu_e,e^-)^{97}$Tc. In this work we follow two different procedures to calculate the contribution to the cross sections coming from the Gamow-Teller and Isobaric Analogue State on one hand and the spin-dipole and higher multipoles (up to $J=5$) on the other hand. In the first one we use the allowed approximation, which corresponds to the low momentum
transfer, as done in Haxton and Johnson's paper. For all other multipoles
we use the Walecka formalism \cite{Walecka}
for the reaction cross section without neglecting the momentum transfer, with the transition matrix elements provided by
the microscopic Quasi-Particle Random-Phase Approximation (QRPA).
Note that this is at variance with \cite{Haxton:1987bf} where only the
spin-dipole is included whose contribution is estimated using the
Goldhaber-Teller model.

\begin{figure}[hbt]
\centering
\includegraphics[width=0.8\textwidth]{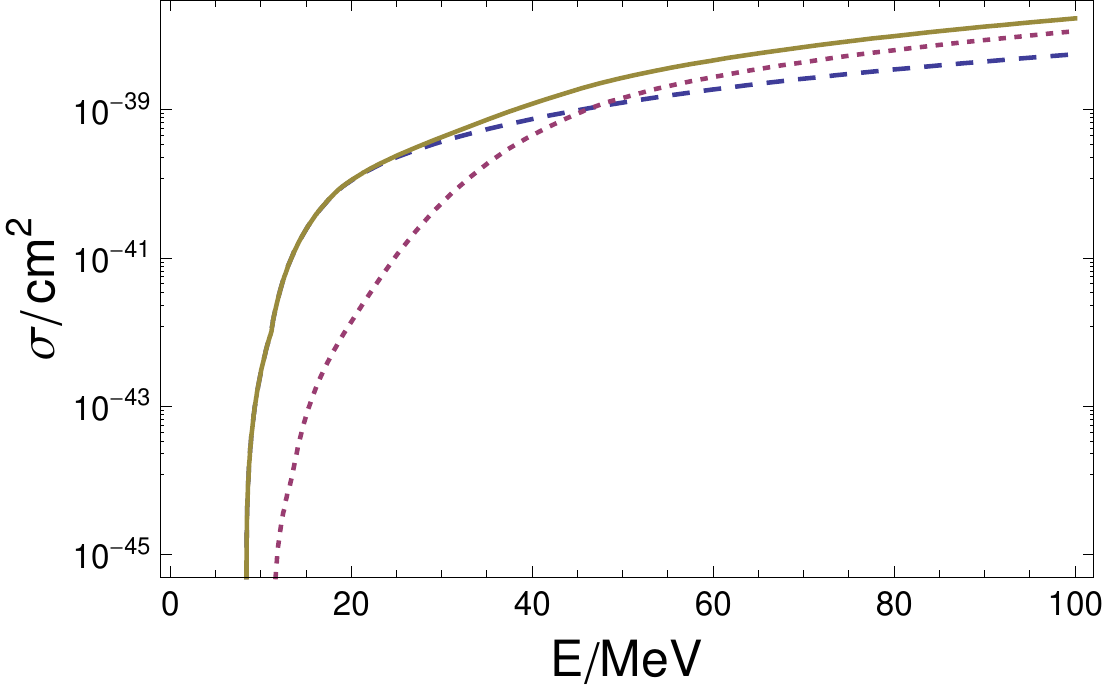}
\caption{\label{fig:cross} ${\rm ^{98}Mo(\nu_e,e^-n)^{97}Tc}$ reaction cross
section as a function of the neutrino energy. The dashed curve presents the
contribution due to the Gamow-Teller and Isobaric Analogue State
transitions, evaluated within the allowed approximation using experimental
data. The dotted curve shows the contribution coming from the spin-dipole and
other multipoles (up to $J=5$), calculated with the QRPA approach.
The total cross section is also shown (solid line).}
\end{figure}

To evaluate the transition matrix elements of the Gamow-Teller and
Isobaric Analogue State within the allowed approximation,
we exploit the only measurement available, the one
by J. Rapaport et al.~\cite{Rappaport}, who
studied $^{98}$Mo$(p,n)^{98}$Tc reaction at forward angles for 120 MeV
incident protons.
This procedure is the same as in \cite{Haxton:1987bf}.
It is important to remind that the model-independent Fermi and Ikeda 
sum-rules tell us that the total strength associated to the Fermi and
Gamow-Teller transitions is given by N-Z and 3(N-Z) respectively,
N and Z being the number of neutrons and protons of the considered nucleus.
In \cite{Rappaport} it is found that
a fraction of the Gamow-Teller strength B(GT), i.e. $\sum $B(GT)=28 $\pm$ 5,
is distributed to excite the $^{98}$Tc states having
up to 18 MeV above the ground state. Since the experimental data are cut
above 18 MeV we have checked the influence of a B(GT) strength located at
higher energy by extrapolating the experimental data with different curves
(Gaussian or Lorentzian).
The results (sec. \ref{sec:results}) turn out not to be sensitive to the specific extrapolation curve chosen.
On the other hand, the Fermi transitions are dominated by
the Isobaric Analog State of $^{98}$Mo,
located at $E=9.656$ MeV above $^{98}$Tc ground state.
This single
resonance carries all the Fermi strength B(F) given by the Fermi sum-rule, namely
B(F) = (N-Z) = 14. Concerning the forbidden transitions,
their contribution to the cross section has been calculated using the QRPA
approach. The largest contribution comes
from the $1^{-}$ and $2^{-}$ transitions to $^{98}$Tc
which essentially affects the high energy part of the cross section (above 40 MeV).
The total cross section we use in this work is shown in Figure \ref{fig:cross}.

The need to minimize the background from solar \ns\ motivates the choice of the process $^{98}$Mo$(\nu_e,e^-n)^{97}$Tc, where one neutron is emitted. For this reaction the  threshold in energy for neutron emission is high enough (8.96  MeV) to suppress the pp solar \n\ flux completely and the $^8B$ and $hep$ fluxes partially compared to the \sn\ contribution. 
We make the assumption that all the excited states of $^{98}$Tc laying above neutron
emission threshold will decay by emitting neutrons and
produce $^{97}$Tc in its ground state.
Note that the $^{97}$Mo+p threshold is by $1.1$ MeV lower than $^{97}$Tc+n one.
However proton emission at
low energies should be highly suppressed compared to the neutron one due to the Coulomb barrier.
At higher energies  the $^{97}$Mo+p channel should become as much
important as the $^{97}$Tc+n one. Thus, at these energies, our predictions
should overestimate the rate of $^{97}$Tc production from $^{98}$Mo.

A delicate task is to accurately predict the $^{97}$Mo$(\nu _e,e^{-})^{97}$Tc
cross section, since no experimental information is available.
Following ref~\cite{Haxton:1987bf} for this case
we have evaluated the $^{97}$Mo$(\nu_e,e^{-})^{97}$Tc cross section exploiting
the sum-rule argument, namely scaling the B(GT)
strength from the $^{98}$Mo$(p,n)^{98}$Tc measurement by the ratio
(N-Z)$_{^{97}{\rm Mo}}/$(N-Z)$_{^{98} \rm Mo}$. This scaling is performed
after having subtracted the near-threshold events describing the transition
to to the $^{98}$Tc ground state. In fact, the
$^{97}$Mo$(\nu_e,e^{-})^{97}$Tc$_{g.s.}$ transition matrix elements is negligibly small
as one can check by converting the $^{97}$Tc$_{g.s.}$ lifetime
($\tau _{1/2}=2.6\cdot 10^{6}s$).
Concerning the Fermi strength, we consider that all the strength
B(F)=2T=13 is associated to the Isobaric Analogue State located at
$11.05$ MeV excitation energy of the $^{97}$Tc nucleus.
As in the $^{98}$Mo case, we neglect $^{97}$Tc production above the $^{96}$Tc+n threshold.
Since again we make the assumption that 100$\%$ of the cases a neutron is
produced, this time we slightly underestimate the $^{97}$Tc production from $^{97}$Mo, which
will give in particular fewer \sne\ events.

%%%%%%%%%%%%%%%%%%%%%%%%%%%%%%%%%%%%%%%%%%%%%%%%%%%%%%%%%%%%%%%%%%%
\section{Results}
\label{sec:results}

\subsection{Rates}

\begin{figure}[hbt]
\centering
\includegraphics[width=0.95\textwidth]{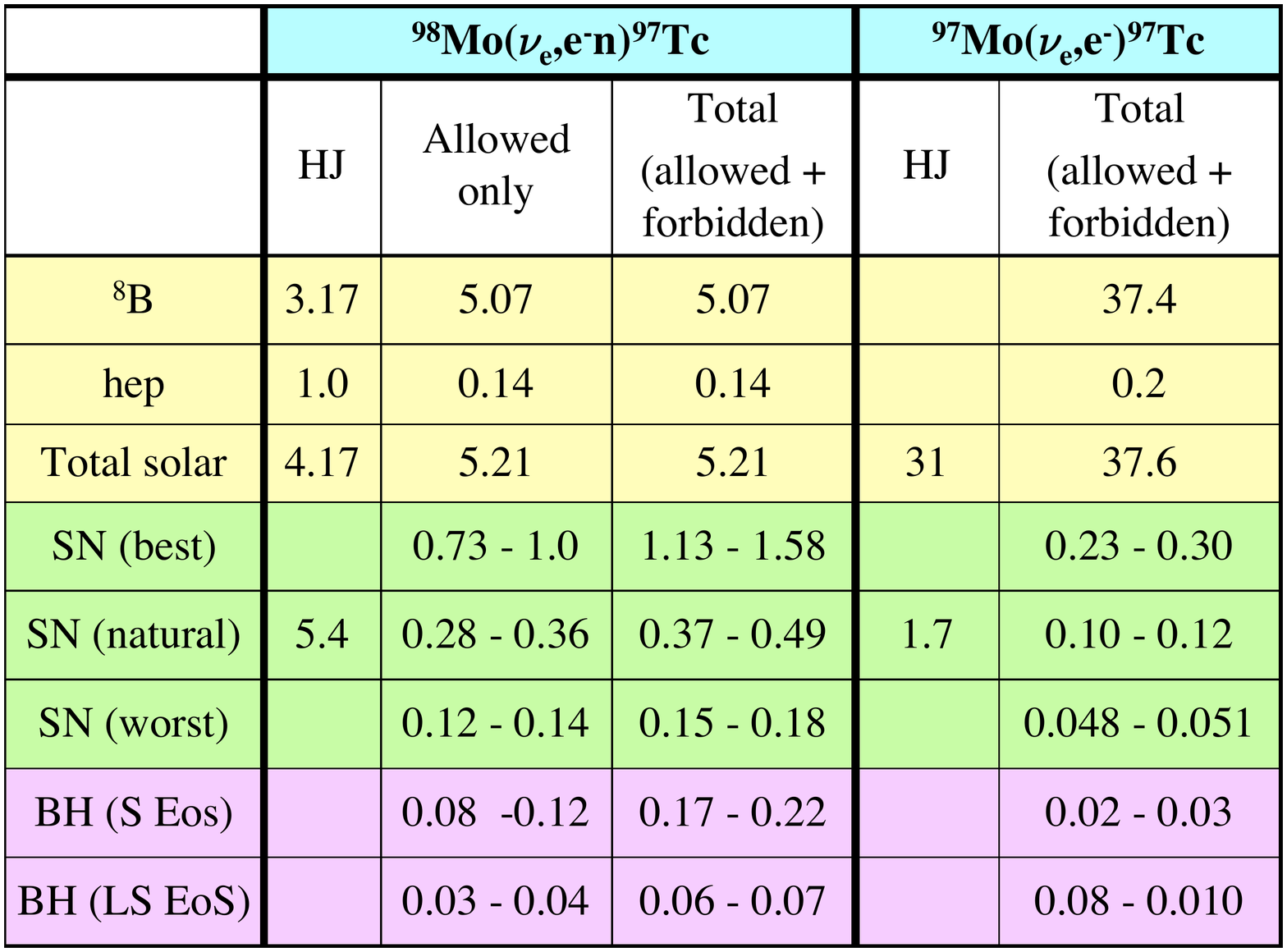}
\caption{\label{table_big} $^{97}{\rm Tc}$ production rates ($10^{-37}
\times s^{-1}$) associated to \ns\ from the sun and supernovae (where
BH indicates failed, black hole-forming \sne), in comparison with
Haxton and Johnson's (HJ) ones.  Intervals for SN rates take into
account the variations in the neutrino oscillation pattern (normal vs
inverted mass hierarchy and small vs large $\protect\theta_{13}$%
). All our results are obtained with a galactic rate  $R_{sn}=0.01~ {\rm yr
^{-1}}$ ($R_{bh}=0.001 {\rm yr
^{-1}}$) for successful (failed) \sne. They can be easily rescaled for other
values of this quantity.  Natural values of $R_{sn}$ are
$0.02 - 0.03$ yr$^{-1}$, but the larger interval $0.01 - 0.05 $
yr$^{-1}$ is allowed (table \protect\ref{tab:rate}). HJ's numbers refer to a supernova rate of
0.09 yr$^{-1}$. }
\end{figure}

The table in figure \ref{table_big} presents our predicted $^{97}$Tc production
rates from $^{98}{\rm Mo}$ and $^{97}{\rm Mo}$ due to both solar and supernova neutrinos.

Let us first comment on the ${\rm ^{98}Mo(\nu_e,e^-n)^{97}Tc}$ channel.
Solar neutrinos produce $^{97}$Tc essentially via the Gamow-Teller and the
Fermi transitions to excited states of $^{98}$Tc, which then decay by emitting
a neutron; the contribution from forbidden transitions is negligible.
In comparison to HJ \cite{Haxton:1987bf}, we find a higher rate of transition due to ${\rm ^8B}$ \ns, 
result of the larger (almost double) normalization of this flux after oscillations (sec. \ref{sec:fluxes}). 
In contrary,  the rate due to the $hep$ flux is an order of magnitude smaller
because of the smaller initial normalization and the further reduction of the
flux due to flavor conversion.

In Figure \ref{table_big} we show predictions for  supernova
neutrinos considering a galactic supernova rate of 0.01 yr$^{-1}$ and taking
into account a range of values for the supernova neutrino fluxes at
the neutrino-sphere and at the surface of the supernova, due to the
unknown neutrino mixing parameters (Table \ref{table:inputs}, fig. \ref{fig:fluxes}).  Contrary to the solar case, the
rate from supernova neutrinos has an important contribution ($\sim$30-60\%) coming
from the forbidden transitions since the neutrino spectra cover
several tens of MeV energy range.  Obviously, an increased luminosity or average neutrino
energy boosts the $^{97}$Tc production rate.  One can see that our
predictions are typically one order of magnitude smaller than what
Haxton and Johnson estimated. The difference is mostly due 
to a higher \sn\ rate, $R_{sn}=0.09$ yr$^{-1}$ and a higher \n\ luminosity ($8 \cdot 10^{52}~{\rm ergs}$ compared to our $L_0=0.5 \cdot 10^{53}~{\rm ergs}$), 
as well as  the fact that the flavor permutation of \sn\ \ns\ was not included by HJ.  Inclusion of oscillations increases the \sn\ flux above $\sim 15-20 $ MeV, as observed in sec. \ref{sec:fluxes}, and therefore lessens the gap between our results and those in \cite{Haxton:1987bf} \footnote{If we make the same assumptions as in
\cite{Haxton:1987bf} we get the same solar background (with differences up to 8$\%$) and a
supernova rate about 17$\%$ lower. This is probably due to the
different procedure followed in the present work as far as the
forbidden transitions are concerned.}.
Still,  while HJ concluded  that the rate of  $^{98}$Mo$(\nu_e,e^-n)^{97}$Tc reaction due to \sn\ \ns\  should
be as much as 1.3 times the solar one, our finding is that the \sn\ contribution is most likely at the level of tens of per cent, but can amount to more than half the solar contribution accepting optimistic \n\ parameters and $R_{sn}\gta 0.03~{\rm yr^{-1}}$, as allowed by current estimates (table \ref{tab:rate}).

The effect of failed \sne\ on the $^{98}$Mo$(\nu_e,e^-n)^{97}$Tc  rate is significant.  While certainly small compared to the solar rate, for the S EoS this contribution can easily  reach $\sim 50\%$ of the one from successful \sne, and can be even of the same size for the largest fraction of failed versus successful \sne, $R_{bh}/R_{sn}\sim 0.2$.  For the LS EoS the effect is more modest, but still at the level of $\sim 10-30\%$ of the total \sn\ contribution. We notice that, due to the more energetic \n\ fluxes, the rate from failed \sne\ is very sensitive to the forbidden states, and nearly doubles when these are included in the cross section calculation. 

The $^{97}$Tc production via the $^{97}$Mo$(\nu_e,e^-)^{97}$Tc channel
constitutes a very different case, also illustrated in
fig. \ref{table_big}.  For this process the reaction threshold is only
320 keV, and therefore a large fraction of the solar $^8$B and $hep$
neutrino fluxes contribute to it. Moreover at 9.8 MeV the
$^{97}$Mo$(\nu_e,e^-n)^{96}$Tc channel opens and starts dominating
over the $^{97}{\rm Tc}$ production, further suppressing the \sn\
contribution to this channel.  For these reasons the $^{97}$Tc
production from $^{97}$Mo is largely dominated by solar neutrinos, primarily due to the
GT transitions, with the \sn-induced production being only a few per
cent of the total. The effect of the
$^{97}$Mo$(\nu_e,e^-)^{97}$Tc channel is mostly to dilute the \sn\
signal due to scattering on $^{98}$Mo into a larger background and make a measurement more challenging.

\subsection{Abundances of $^{97}$Tc in Molybdenum rocks}

In order to estimate the feasibility of a geochemical experiment one
should translate the calculated rates into the number of $^{97}$Tc
atoms in a sample of rock, which is the experimentally measurable
quantity.
We calculated this for 10 kt mass of Molybdenum ore, considering that it contains about 49 t of ${\rm MoS_2}$ \cite{Haxton:1987bf}, and assuming that its Mo content follows the natural
isotopic abundances (i.e., 24.13\% for ${\rm ^{98}Mo}$ and 9.55\% for ${\rm ^{97}Mo}$).

\begin{figure}[hbt]
\centering
\includegraphics[width=0.95\textwidth]{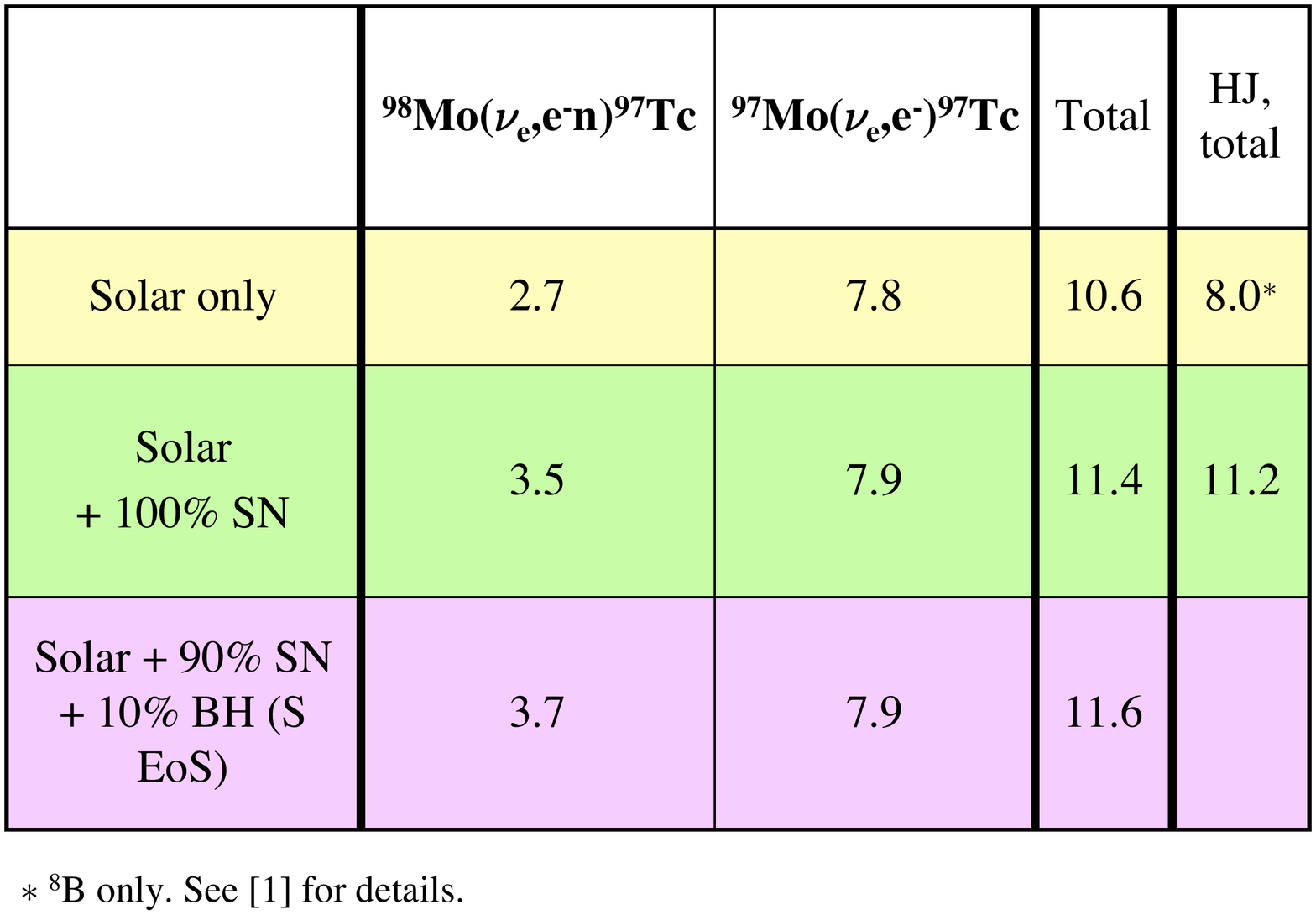}
\caption{\label{table:abundances} Numbers of $^{97}{\rm Tc}$ atoms ($10^6$ unit) expected in 10 kt of
rock from the two different processes of interest.
Totals are given too, as well as the results for the capture rates of Haxton
and Johnson (HJ) for  comparison. Our results
refer to a supernova rate of $0.03~{\rm yr^{-1}}$ and
parameters in the ``natural" scenario (Table \protect\ref{table:inputs}) for successful supernovae (SN).  Results in the last row refer to a scenario with a 10\% failed supernovae (BH) with neutrino emission as in the S model of ref. \protect\cite{Nakazato:2008vj} (see fig. \protect\ref{fig:fluxesbh}). In all cases we have considered complete flavor permutation ($p=0$), which gives the largest ${\rm ^{97}Tc}$ production rates for successful SNe.}
\end{figure}
Results are presented in fig. \ref{table:abundances}. It appears that in 10 kt
of Molybdenum ore that is completely shielded from cosmic rays, there
should be $\simeq 1.1\cdot 10^7$ atoms of ${\rm ^{97}Tc}$, of which
about 10\% from \sne, for natural parameters of \sn\ \n\ emission and
$R_{sn}=3\cdot 10^{-2}~{\rm yr^{-1}}$.  This contribution can be
larger by up to a factor of 3 if the best scenario of \sn\ \n\
spectra and luminosities (table \ref{table:inputs}) is realized. The effect of
failed \sne\ is a few per cent of the total abundance. We note how our
calculated abundance is very close to that of Haxton and Johnson:
this similarity is completely accidental, since our input quantities
are very different and we find a substantially higher solar
contribution, balanced by a lower \sn\ one.

Let us now discuss the significance of our result in view of an
experiment.  The first question is if the neutrino-induced ${\rm
^{97}Tc}$ is at detectable level. The answer is affirmative: already
at the time of HJ the best instruments were quoted
\cite{Haxton:1987bf} to have sensitivity down to about $5 \cdot 10^6$
atoms for a pure Tc sample. A modern experiment would employ essentially the same methods, with a sensitivity slightly improved (a factor of two or so) by a number of technological updates \cite{anbarprivate}.
%
%Today it is estimated \cite{.} that
%.... \ceci{add paragraph from A. Anbar.. I have made a request to him.. waiting to hear back.}.  
%
Therefore we come to the
same conclusion as HJ: at the price of setting up an industrial-scale
extraction project, capable to process about 10 kt of rock at large
depth\footnote{We stress that the use of industrial-scale equipment and large scale excavations -- while exceeding the typical scale of geophysics projects -- are common in the field of neutrino astrophysics.}, an experiment should be feasible.  Such
experiment should be arranged in such a way to avoid the ``roast
memory'' problem that was fatal to the original experimental attempt
in 1988 \cite{amm}. While potentially laborious, this is not difficult conceptually.

Once established that a detection is possible, a key question is if
the supernova neutrino contribution can be unambiguously identified.
Our study makes it clear that this contribution is comparable to the
several uncertainties involved in the problem. Indeed, already the GT
strength distribution obtained from $^{98}$Mo$(p,n)^{98} {\rm Tc}$
measurement has more than 17\% uncertainty, whereas the
$^{97}$Mo$(\nu,e^-)^{97}$Tc cross section estimate along with
branching ratios into the decay channels is purely
phenomenological. One should also consider the $\sim 5-10\%$ error on
the oscillated $^8{\rm B}$ solar \n\ flux, and the still unmeasured
hep flux. 

Therefore, we conclude that a measurement of the $^{97}$Tc abundance in deep Molybdenum ore would be an interesting complement to studies of the solar \n\ flux, but would provide only loose upper limits on the \sn\ flux, perhaps at the level of the constraint from \n\ detectors on the \sn\ rate (table \ref{tab:rate}).
Two major experimental steps should be achieved in the near future to render the geochemical quest for galactic supernova neutrinos attractive : i) an increase in precision in the measured  $^8$B solar \n\ flux and a new precise measurement of the $hep$ flux; ii) a precise meaurement of the
transition matrix elements for the Gamow-Teller as well as the forbidden multipoles contributing to the
$^{97}$Mo$(p,n)^{97}$Tc and $^{98}$Mo$(p,n)^{98}$Tc reactions, along with the branching ratio
relevant to $^{97}$Tc production. This can be obtained from future charge-exchange measurements for example at KVI or RIKEN. A 
direct measurement of the neutrino induced cross sections could also be performed at future low energy neutrino facilities exploiting conventional sources, such as at $\nu$SNS, or at a low energy
beta-beam facility \cite{Volpe:2003fi,Lazauskas:2007bs,Volpe:2006in}.

\section{Conclusions}

In this paper we have investigated the possibility to observe the integrated supernova galactic neutrino
contribution by measuring the $^{97}$Tc produced by electron neutrinos in a Molybdenum ore, as first
proposed in~\cite{Haxton:1987bf}. With this aim realistic estimates -- reflecting the state of
the art in the neutrino physics and of galactic \sne\ neutrino fluxes -- have been performed.
For our calculations we have used experimental information on the Gamow-Teller and Fermi distributions of the $^{98}$Mo$(\nu_e,e^-)^{98}$Tc
reaction, combined with QRPA predictions for the forbidden contributions.

We have found that the relative production of  $^{97}$Tc from \sn\ neutrinos is about 0.1
of the solar background, which is much smaller than what was evaluated in the pioneering study~\cite{Haxton:1987bf}.
For a geochemical measurement to become attractive, a better
knowledge of the solar neutrino spectrum and of the $^{97}$Mo$(\nu_e,e^-)^{97}$Tc and  $^{98}$Mo$(\nu_e,e^-n)^{97}$Tc
cross sections are necessary.  An improved experimental sensitivity is desirable to reduce the quantity of rock to be processed. Future precise charge-exchange measurements
at high energies and forward angles and/or neutrino-nucleus measurements for the $^{97}{\rm Mo}$ and $^{98}{\rm Mo}$ isotopes will defintely be very helpful.

\vspace*{0.5cm}
We are indebted to W. Haxton and A. Anbar for very fruitful exchanges.  C.V. and R.L. acknowledge the support from "Non standard
neutrino  properties and their impact in astrophysics and cosmology", Project No. ANR-05-JCJC-0023.  C.L. acknowledges ASU and RBRC for support, and is grateful to the IPN (Orsay) for generous hospitality while working on this project.

%%%%%%%%%%%%%%%%%%%%%%%%%%%%%%%%%%%
\bibliography{draftJCAP}

\end{document}